\RequirePackage{fix-cm}
\documentclass[twocolumn,epjc3]{svjour3}
\smartqed  
\RequirePackage{graphicx}
\RequirePackage{amsmath,amssymb}
\RequirePackage{booktabs}

\begin{document}

\title{Salvaging Power-Law Inflation through Warming 
}

\author{M. Alhallak\thanksref{e1,addr1},
N. Chamoun\thanksref{e2, addr2}
        \and
        M. S. Eldaher\thanksref{e3,addr3} 
}

\thankstext{e1}{phy.halak@hotmail.com}
\thankstext{e2}{nidal.chamoun@hiast.edu.sy}
\thankstext{e3}{m-saemaldahr@aiu.edu.sy}

\institute{Physics Dept., Damascus University, Damascus, Syria \label{addr1}
           \and
           Physics Dept., HIAST, Damascus, Syria\label{addr2}
           \and
           Higher Institute of Laser Applications and Researches, Damascus University, Damascus, Syria \label{addr3}
}


\maketitle

\begin{abstract}
Power-Law inflation with scale factor $a \propto t^m$ is investigated in the context of warm inflation. The treatment is performed in the weak and strong dissipation limits. In addition, we discuss the three common  cases for the thermal dissipation coefficient $\Gamma(T)$. We compare the theoretical results of the Power-Law model within warm inflation with the observational constraints from Planck $2018$ and BICEP/Keck 2018, as presented  by the tensor-to-scalar ratio $r$ and spectral index $n_s$. The model results agree largely with the observations for most of the $\Gamma(T)$ cases.

 Furthermore, in order to address the problem of exiting the inflationary epoch, we suggest a perturbed modification to the power-law definition so that it becomes affine, and find that this small change indicates a way for having
an exit scenario with a suitable e-foldings number. Finally, we examine this perturbation ansatz within the context of cold inflation with exponential potential, and we find that it can accommodate the observational data with sufficient e-foldings.

Our study suggests that the power-law
inflation and the exponential potential, in both warm
and cold inflation contexts, can in principle be made consistent with the
observations and with a possible graceful exit.
\end{abstract}

\section{Introduction}
\label{intro}
The inflationary paradigm was suggested to sort out several drawbacks of the standard Big Bang theory, such as the horizon, flatness, and monopole problems \cite{Starobinsky:1980te,Guth:1980zm,Sato:1980yn,Linde:1981mu,Albrecht:1982wi,Linde:1983gd}\footnote{It is noteworthy that the purely geometrical model of \cite{Starobinsky:1980te} remains valid and produces an acceptable fitting with current observational data, in contrast to some models proposed later.}. The second benefit of this paradigm is that it produces the observed anisotropy of the cosmic microwave background \cite{Bennett:2010jb,WMAP:2012nax,Meerburg:2013dla,WMAP:2010qai,Larson:2010gs} as well as the distribution of the large-scale structure \cite{Starobinsky_jetp,Mukhanov:1981xt,Hawking:1982cz,Guth:1982ec,Starobinsky:1982ee,Bardeen:1983qw}.

  It is commonly assumed that inflation happens when a single scalar field, the inflaton, be it a matter field or a geometric entity (e.g. \cite{Kallosh}) or expressing a distinct physical concept such as variation of constants (e.g. \cite{JCAP-Chamoun}), slowly rolls down to a nearly flat potential and induces a quasi-de Sitter phase. Typically, the inflaton's quantum fluctuations generate adiabatic density fluctuations. Inflation potentials must be carefully selected; otherwise, they will overproduce density perturbations.
Inflationary models are often restricted by the assumption that the inflaton is only (minimally) coupled to gravity. However, introducing couplings to other early Universe sectors can alleviate these restrictions.

Warm inflation \cite{Berera:1995ie,Berera:1995wh}, a well-established alternative to conventional (cold) inflation, entails thermally coupling an inflaton field to a radiation bath. The fluctuations in warm inflation are mainly thermal, with quantum fluctuations being suppressed when dissipation rates between the inflaton field and radiation sector are large. In contrast to cold inflation, the inflaton produces radiation continuously, eliminating the necessity for a reheating phase at the end of inflation.

Two different perspectives can be taken into account when analyzing inflationary scenarios. As a first approach, we choose a scalar field's potential, which plays a key role in the inflation scenario construction. This type of method is known as the ``potential motivated approach”. Therefore, inflationary models are classified, within this approach, according to their scalar field potentials. In contrast, one can, on the other hand, reconstruct the potential $V$ from the dynamics of the expansion, which are determined by a scale factor $a(t)$. As a result of this approach, which we shall call a ``dynamically motivated approach", inflationary models are categorized based on the corresponding universe expansion laws.

In this paper we consider an example of the second approach, power-law inflation \cite{Lucchin:1984yf}, where the scale factor depends on time as $a=a_0 t^m$.
Several studies have examined the power-law expansion in the context of cold inflation \cite{Sahni:1988zb,Sahni:1990tx,Halliwell:1986ja,Burd:1988ss,Malik:1998gy,Copeland:1999cs}.
There were, however, two main problems with this model. First, the scalar-to-tensor ratio $r$ turns out to be larger than the limits set by the Planck data, and the model, thus, can be ruled out. Second, the model has no exit scenario because the parameters for the slow roll are constant.

There have been several attempts to improve the model's results in the context of cold inflation  \cite{Tsujikawa:2000tm,Unnikrishnan:2013vga}. Some studies \cite{Yokoyama1988,Liddle1989} accounted also for the possibility of thermal effects of viscosity caused by couplings of the inflaton and other particles, exhibiting explicitly dissipation terms in the inflation equations, expressing decay of inflaton into radiation, which can be considered as reminiscent of warm inflation. Warm inflation with exponential potential were recently studied in \cite{1910.06796,2007.15268} in the context of landscape/swampland of string theory, albeit only for $T$-cubic dissipation term, where the effects of the braneworld extra dimensions allow achieving a strong dissipative regime.


The purpose of this letter is to present a general discussion of power law inflation within  a warm context. Thus, we shall investigate the early power-law expansion of the universe within the warm inflation paradigm, considering several forms for the  dissipation coefficients, and studying the scalar fields in the weak and strong dissipation limits to radiation. Furthermore, in order to find a graceful exit way, we introduce an affine perturbation on the power-law scale factor, and find that its repercussion on the potential, albeit tiny, is hinting for a way to provide for a successful exit from the inflationary epoch. Moreover, when assumed within the cold power-law inflation, we show that our perturbing modification can also improve it.

The paper is organized as follows. In section 2, we summarize the main findings of the warm inflation, considering three different forms of dissipation parameter $\Gamma$. We then present formulas corresponding to the main observables: spectral index $n_s$ and tensor-scalar ratio $r$ under the weak and strong dissipation limits. Section 3 analyzes the power-law model within the context of warm inflation. After computing the potentials of different $\Gamma$ cases and dissipation limits, the observational results of the model are compared to those of Planck 2018 and BICEP/Keck 2018. In section 4, we present a possible perturbative exit scenario in order to accomplish the end of inflation, and apply it also in the context of cold inflation. We end up with a summary and conclusion in section 5.

\section{Warm Inflation }
\label{sec:Warm Inflation}
The main results of warm inflation will be reviewed in this section, focusing on strong and weak dissipation limits. Warm inflation is characterized by the decay of the inflaton field into radiation during the inflationary epoch. Scalar field equations and radiation equations both express this decay as,
\begin{equation}
\label{eq:EOM_rad}
\dot{\rho}_{R}+4H\rho_R=\Gamma \dot{\phi}^2
\end{equation}
\begin{equation}
\label{eq:EOM_phi}
\dot{\rho}_{\phi}+3H(\rho_\phi + P_\phi) =-\Gamma \dot{\phi}^2
\end{equation}
The energy density $\rho_\phi$ and pressure $P_\phi$ of the inflaton field can be expressed in canonical form as,
\begin{equation}
\rho_\phi=\frac{1}{2}\dot{\phi}^2+V(\phi) ,  P_\phi=\frac{1}{2}\dot{\phi}^2-V(\phi)
\end{equation}
On the other hand, the energy density $\rho_R$ and the pressure $P_R$ of radiation are given as,
\begin{equation}
\label{eq:rho_R}
\rho_R=\alpha T^4 ,  P_R=\rho_R/3
\end{equation}
with $\alpha=\pi^2/30 g(T)$ ($g(T)$ is related to the number of massless modes such that $\alpha \approx 75$ \cite{Visinelli:2011jy}). This corresponds to a number of degrees of freedom $g=228.75$, i.e. of the minimal supersymmetric (susy) standard model (MSSM). This is a plausible choice as the energy scale of the inflationary period is commonly taken to be around the GUT scale $\sim 10^{16}$ GeV. Assuming the breaking of the underlying GUT (rather, its susy version, which has some benefits compared to the non-susy version, such as accommodating in an easier way the gauge couplings unification) occurs at one stage enforcing a big desert picture in that no other particles are created up to that scale, then the number of relativistic particles with which the inflaton will thermalize would be that of the theory governing the low energy scale which, for simplicity, will be that of the MSSM embedded in most susy GUTs.

In Einstein's gravity with minimal coupling between the scalar field and the gravitational sector, Friedmann equation has the form,
\begin{equation}
\label{eq:Freidmann}
3 H^2= \rho_{tot}= \big(\frac{1}{2}\dot{\phi}^2+V(\phi)+\rho_R\big)
\end{equation}
Here, and henceforth, we work in units where the reduced Planck mass is set to one.

\subsection{ Slow-roll regime of warm inflation}
Assuming the potential energy dominates all other forms of energy, the slow-roll approximation is expressed as,
\begin{equation}
\ddot{\phi}\ll H \dot{\phi},\qquad \dot{\rho_R}\ll H \rho_R,\qquad \frac{1}{2}\dot{\phi}^2 \ll V(\phi)
\end{equation}
As a result, Eqs,(\ref{eq:EOM_rad}),(\ref{eq:EOM_phi}) and (\ref{eq:Freidmann}) are read as,
\begin{equation}
\label{eq:rho_R_SRA}
\rho_R \simeq \frac{3}{4} Q  \dot{\phi}^2
\end{equation}
\begin{equation}
\label{eq:dotphi_SRA}
\dot{\phi} \simeq -\frac{V_\phi}{3H(1+Q)}
\end{equation}
\begin{equation}
\label{H_SRA}
3 H^2 \simeq V
\end{equation}
where,
\begin{equation}
\label{Q}
Q \equiv \frac{\Gamma}{ 3 H}
\end{equation}
represents the effectiveness at which the inflaton converts into radiation.

We can find the slow-roll parameters as,
\begin{equation}
\varepsilon= \frac{1}{2} \bigg(\frac{V_\phi}{V}\bigg)^2 , \eta \equiv \frac{V_{\phi\phi}}{V},\beta \equiv \frac{\Gamma_\phi V_\phi}{\Gamma V}
\end{equation}
As opposed to the cold inflation, the slow-roll condition appears now in the warm inflation as:
\begin{equation}
\label{slowrollconditions}
    \frac{\varepsilon} {(1+ Q)} \ll 1,
    \frac{\eta} {(1+ Q)} \ll 1,
    \frac{\beta} {(1+ Q)} \ll 1.
\end{equation}

\subsection{ perturbations spectra}
The primordial power spectrum of warm inflation at the horizon crossing is given by \cite{Bastero-Gil:2016qru} ,
\begin{equation}
\Delta_R^2= A_s \big(1+ 2 n_\ast + \omega_\ast \big)G(Q) : A_s = \frac{V_\ast (1+Q_\ast)^2}{24 \pi^2 \varepsilon_\ast}
\end{equation}
where $\ast$ denotes the parameters at the horizon crossing,  $n=\big(\exp H/T -1 \big)^{-1}$
is the Bose-Einstein statistical function, $\omega=\frac{T}{H}\frac{2\sqrt{3}\pi Q}{\sqrt{3+4\pi Q}}$  and $A_s$ represents the amplitude of the CMB fluctuations. The function $G(Q)$ describes the growth of inflaton fluctuations due to the coupling to radiation.

In our study, we shall investigate three different cases of the dissipation parameter ($\Gamma (T)$). The function $G(Q)$ is determined for each case \cite{Benetti:2016jhf} :
\begin{eqnarray}
\mbox{First dissipation case: constant parameter} \nonumber \\
 G(Q) = 1, \mbox{for } \Gamma = \Gamma_0 = constant,
\end{eqnarray}
\begin{eqnarray}
\mbox{Second dissipation case: linearly $T$-dependent parameter} \nonumber \\
G(Q) = 1+0.335 Q^{1.364}+0.0185 Q^{2.315}, \mbox{for } \Gamma = \Gamma_0 T,\nonumber\\
\end{eqnarray}
\begin{eqnarray}
\mbox{Third dissipation case: cubically $T$-dependent parameter} \nonumber \\
G(Q) = 1+4.981 Q^{1.946}+0.127 Q^{4.330}, \mbox{for } \Gamma = \Gamma_0 T^3. \nonumber\\
\end{eqnarray}

The general formula for temperature $T$ can be found by combining Eqs (\ref{eq:rho_R}), (\ref{eq:rho_R_SRA}),(\ref{eq:dotphi_SRA}) and (\ref{H_SRA}) in the slow-roll approximation as,
\begin{equation}
T=\bigg(\frac{V}{2\alpha}\frac{Q_\ast}{(1+Q_\ast)^2}\varepsilon_\ast \bigg)^{1/4}
\end{equation}
The scalar spectral index is determined by,
\begin{equation}
n_s=\lim_{k -> k_\ast}\frac{d \ln \Delta_R^2(k/k_\ast)}{d \ln(k/k_\ast)}
\end{equation}
The tensor-to-scalar perturbation ratio, $r$ is,
\begin{equation}
r=\frac{\Delta_T^2}{\Delta_R^2}
\end{equation}
where $\Delta_T^2$ is the power spectrum of the tensor perturbation, \begin{equation}
\Delta_T^2=2 H^2/\pi^2
\end{equation}

\subsubsection{Weak Dissipation Limit}
In this section, we will cover the principal formulas of the observables $n_s$ and $r$ in the weak dissipation limit $Q \ll 1$. It is crucial to note the differences in the formulas for the spectral index $n_s$ as well as the temperature $T$ formulas based on $\Gamma(T)$. We summarize these formulas below:
\begin{equation}
\label{eq:Tweak}
T = \bigg(\frac{V_\phi^2 \Gamma_0}{36 \alpha H^3} \bigg)^{x}
\end{equation}
\begin{equation}
\label{eq:ns_weak}
n_s-1 = (-6 \varepsilon + 2\eta)+\frac{2\pi \Gamma T}{12 H^2}(15 \varepsilon - 2 \eta) + y Q^z (\varepsilon-\beta  )
\end{equation}
\begin{equation}
\label{eq:r_weak}
r=\frac{H}{2 T}16 \varepsilon
\end{equation}
where:
\\for $\Gamma=\Gamma_0$
\begin{equation}
x=\frac{1}{4}  ,\qquad y=0 ,\qquad z=0
\end{equation}
for $\Gamma=\Gamma_0 T$
\begin{equation}
x=\frac{1}{3}  ,\qquad y=0.456 ,\qquad z=1.364
\end{equation}
for $\Gamma=\Gamma_0 T^3$
\begin{equation}
\label{lastofweak}
x=1  ,\qquad y=9.69 ,\qquad z=1.946
\end{equation}

\subsubsection{Strong Dissipation Limit}
Similarly, in the strong dissipation limit where $Q\gg1$,
we can find the following relations:
\begin{equation}
\label{eq:Tstrong}
T = \bigg(\frac{V_\phi^2 }{4 \alpha \Gamma_0 H} \bigg)^{x}
\end{equation}
\begin{equation}
n_s-1=\frac{1}{Q}(-\frac{9}{4}\varepsilon+\frac{3}{2}\eta-\frac{9}{4}\beta) - \frac{y}{Q}(\beta-\varepsilon)
\end{equation}
\begin{equation}
r=\frac{H}{T}(\frac{16\varepsilon}{\sqrt{3 \pi} (z Q^{w})Q^{5/2}})
\end{equation}
where:
\\for $\Gamma=\Gamma_0$
\begin{equation}
x=\frac{1}{4}  ,\qquad y=0.0,\qquad z=1,\quad\qquad w=0
\end{equation}
for $\Gamma=\Gamma_0 T$
\begin{equation}
x=\frac{1}{5}  ,\qquad y=2.3,\qquad z=0.0185,\quad w=2.315
\end{equation}
for $\Gamma=\Gamma_0 T^3$
\begin{equation}
x=\frac{1}{7}  ,\qquad y=4.33,\qquad z=0.127,\quad w=4.330
\end{equation}

\section{Warming the Power-Law Inflation}
\label{sec:Warming Power-Law Inflation}
This section will examine the power-law inflation within a warm context. Power-law inflation is characterized by Hubble parameter $H=\frac{\dot{a}}{a}$ given by:
 \begin{equation}
\label{eq:PowerLaw_H}
H=\frac{m}{t}
\end{equation}
where $m \geq 1$ in order to achieve an accelerated inflationary phase. Using Eqs (\ref{H_SRA}) and (\ref{eq:PowerLaw_H}), we  find,
\begin{equation}
\label{eq:Vintermsoft}
V=\frac{3 m^2}{t^2}
\end{equation}

As said earlier, we shall consider the weak dissipation limit $Q\ll 1$, as well as  the strong dissipation one $Q\gg 1$, and our goal is to identify the potential function responsible for this type of expansion, then calculate the crucial observables, represented by the tensor to scalar ratio $r$ and the spectral index $n_s$, in order  to compare with the experimental results of Planck 2018 and BK18.

\subsection{Power-Law inflation in the Weak Dissipation Limit}\label{sect:PoweWeak}
We start by determining the potential function imposing the power-law expansion (\ref{eq:PowerLaw_H}).

Substituting Eq.(\ref{eq:Vintermsoft}) into the integration of Eq.(\ref{eq:dotphi_SRA}), under the weak dissipation limit, we find,
\begin{equation}
\label{eq:phi_weak}
\phi=\sqrt{2} \sqrt{m} \log (t)
\end{equation}
As a result, the potential $V$ is given in terms of the field $\phi$ as,
\begin{equation}
\label{eq:VintermsOfphi_weak}
V= V_0  e^{-\frac{\sqrt{2} {\phi}}{\sqrt{m}}}
\end{equation}
    with $V_0=3 m^2$.

The potential (\ref{eq:VintermsOfphi_weak}) is identical to the case of cold inflation, mainly due to the weak dissipation limit in which $\Gamma \ll H$. This exponential potential can be induced in many contexts, such as a temporal variation of coupling constants, eventually with nonminimal coupling to gravity and/or modified gravity \cite{ijmpaChamoun,universeChamoun}. However, the empirical observables of $n_s$ and $r$ derived from this potential under warm inflation differ dramatically from those under cold inflationary conditions.

   Table (\ref{table:weak_formuals}) summarizes the main formulas of the observables $n_s$ and $r$, along with key characteristic quantities of warm power-law inflation, such as $Q$, $T$, and $H$, following Eqs (\ref{eq:Tweak})-(\ref{lastofweak}).

 \begin{table*}[ht]
\caption{Main specific characteristics of the warm model in the weak dissipative regime with exponential potential. }\label{table:weak_formuals}
\centering
\begin{tabular}{p{0.1\linewidth}p{0.2\linewidth}p{0.35\linewidth}p{0.35\linewidth}}
\hline
& $\Gamma=\Gamma_0$ &  $\Gamma=\Gamma_0 T$ & $\Gamma=\Gamma_0 T^3$\\
\hline
$\varepsilon$ &$\frac{1}{m}$  &  $\frac{1}{m}$ & $\frac{1}{m}$\\
$\eta$ &$\frac{2}{m}$  &  $\frac{2}{m}$ &$\frac{2}{m}$ \\
$\beta$ &$0$ & $\frac{1}{3 m}$ & $\frac{3}{m}$  \\
$Q$ & $\frac{{\Gamma_0} }{3 m} e^{\frac{{\phi}}{ \sqrt{2 m}}}$ &$\frac{{\Gamma_0}^{4/3} }{3 \sqrt[3]{{2 \alpha}} m} e^{\frac{\sqrt{2} {\phi}}{3 \sqrt{m}}}$   & $\frac{{\Gamma_0}^4 }{24 {\alpha}^3 m}e^{-\frac{\sqrt{2} {\phi}}{\sqrt{m}}}$ \\
$T$ &$\frac{\sqrt[4]{{\Gamma_0}} }{ \sqrt[4]{{2 \alpha}}} e^{-\frac{{\phi}}{4 \sqrt{2 m}}}$ &$\frac{\sqrt[3]{{\Gamma_0}} }{ \sqrt[3]{{2\alpha}}} e^{-\frac{{\phi}}{3  \sqrt{2 m}}}$ &$\frac{{\Gamma_0} }{2 {\alpha}}e^{-\frac{{\phi}}{\sqrt{2} \sqrt{m}}}$\\
$n_s-1$ & $\frac{11 \pi {\Gamma_0}^{5/4} e^{\frac{7 {\phi}}{4  \sqrt{2 m}}}}{6 \sqrt[4]{{2 \alpha}} m^3}-\frac{2}{m}$ &$\frac{2.63 {\Gamma_0}^{5/3} e^{\frac{0.94 {\phi}}{\sqrt{m}}}}{{\alpha}^{0.66} m^{3}}+\frac{0.049 {\Gamma_0}^{1.81} e^{\frac{0.64 \phi}{\sqrt{m}}}}{{\alpha}^{0.45} m^{2.36}}-\frac{2}{m}$
&
$-\frac{0.039 {\Gamma_0}^{7.78} e^{-\frac{2.75 {\phi}}{\sqrt{m}}}}{{\alpha}^{5.83} m^{2.94}}+\frac{0.52 {\Gamma_0}^5 e^{-\frac{1.41 {\phi}}{\sqrt{m}}}}{{\alpha}^{4.} m^{3}}-\frac{2}{m}$
\\
$r$ & $\frac{8 \sqrt[4]{{2\alpha}} }{\sqrt[4]{{\Gamma_0}}} e^{-\frac{3 {\phi}}{4 \sqrt{2 m}}}$ & $\frac{8  \sqrt[3]{{2 \alpha}} }{\sqrt[3]{{\Gamma_0}}}e^{-\frac{\sqrt{2} {\phi}}{3 \sqrt{m}}}$  &$\frac{16 {\alpha}}{{\Gamma_0}}$ \\
\hline
\end{tabular}
\end{table*}
By examining table (\ref{table:weak_formuals}), we can see that we have different results from the cold inflation scenario, where the observables $r$ and $n_s$ are determined by equations:
\begin{equation}
\label{eq:nsCold}
n_s-1=\frac{-2}{m}
\end{equation}
\begin{equation}
\label{eq:rCold}
r=\frac{16}{m}
\end{equation}
It is apparent that, unlike the case of cold inflation, the observables may give various and different results  according to the expression of $\Gamma$ as a function of $T$ (our setup), and  of $\phi$ (in other possible setups) as well.

\begin{figure}
  \includegraphics[width=0.5\textwidth]{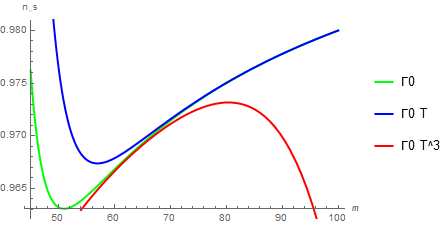}
\caption{Plot of a spectral index $n_s$ as a function of $m$. The green line indicates the $\Gamma=\Gamma_0$ case with ($\Gamma_0=1\times10^{-8}$, $\phi_\ast=163$), the blue line represents the $\Gamma=\Gamma_0 T$  case with ($\Gamma_0=2\times10^{-4}$, $\phi_\ast=178$), and the red line refers to $\Gamma=\Gamma_0 T^3$  with ($\Gamma_0=79\times10^{3}$, $\phi_\ast=200$). }
\label{fig:ns_m_weak}       
\end{figure}

A comparison between $\Gamma=\Gamma_0$, $\Gamma=\Gamma_0 T$ and $\Gamma=\Gamma_0 T^3$ cases, shows that when the third case is considered, the scalar-to-tensor ratio $r$ is, for fixed $\alpha$, independent of the scalar field $\phi$ but dependent only on the parameter $\Gamma_0$, as compared to the other two cases where $r$ is dependent upon all model parameters.
On the other hand, the spectral index $n_s$ has a minimum value in both cases of $\Gamma=\Gamma_0$ and $\Gamma=\Gamma_0 T$, while it has a maximum value in the case of $\Gamma=\Gamma_0 T^3$, as shown in Fig. (\ref{fig:ns_m_weak}) depicting the spectral index for a range of $m \in [45,100]$.

A quick glance at table (\ref{table:weak_formuals}) reveals that the slow-roll parameters are independent of the scalar field $\phi$. Therefore, warm power-law inflation is like the cold one in that it has no end.

\begin{figure*}
  \includegraphics[width=1.\textwidth]{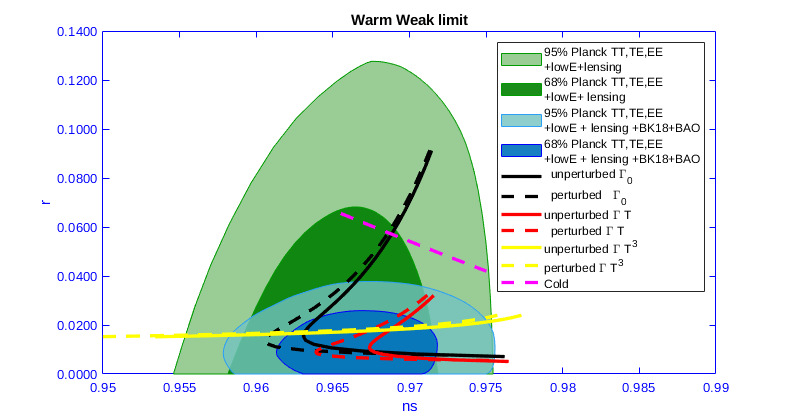}
\caption{Contour plots of $n_s$ and $r$ in the case of weak dissipation limit. Solid lines represent the theoretical results of the model for an exponential potential (Eq. \ref{eq:VintermsOfphi_weak}). A solid black line corresponds to $\Gamma=\Gamma_0$ with $\Gamma_0=1\times10^{-8}$, $\phi_\ast=163$ and $m \in [45-70]$; a solid red line represents $\Gamma=\Gamma_0 T$ with $\Gamma_0=2\times 10^{-4}$,$\phi_\ast=178$ and $m \in [50-70]$. Finally, the yellow solid line indicates the $\Gamma=\Gamma_0 T^3$ case with $m=1\times 10^{2}$, $\phi_\ast=200$ and $\Gamma_0 \in [50-79]\times 10^3$. On the other hand, dashed lines represent the observables resulting from the potential of (Eq. \ref{eq:U}) taken with the same case of $\Gamma$ in their same-color solid lines, with the perturbation parameter $\zeta=-1\times10^{-7}$ ($\zeta=-1.5\times10^{-6}$) for black  and red (yellow) dashed lines. The other parameters for the dashed lines are the same as those for their same-color solid lines.}
\label{fig:ns_r_weak_warm}       
\end{figure*}

In figure (\ref{fig:ns_r_weak_warm}), we compare the observational results of the model with those of Planck $TT-TE-EE+LowE+lensing$ and $BK18$, and see that there is a significant agreement in that the model predictions fall within the $68\%$ confidence level (CL) region for the combined Planck $TT-TE-EE+LowE+lensing \& BK18$.

Table (\ref{table:points_weak}) presents specific benchmarks ($\Gamma_0, m, \phi_*$) giving acceptable values for observables ($n_s$ and $r$), along with the corresponding values of some characteristic model quantities ($Q$, $T$, and $H$) in order to demonstrate that all warm inflationary constraints are met.
\begin{table*}[ht]
\caption{Benchmark points in the case of weak dissipation limit for different instances of dissipation coefficient $\Gamma$, with $\phi_\ast=163$ for $\Gamma=\Gamma_0$, $\phi_\ast=178$ for $\Gamma=\Gamma_0 T$ and $\phi_\ast=200$ for $\Gamma=\Gamma_0 T^3$.  }\label{table:points_weak}
\centering
\begin{tabular}{p{0.08\linewidth}p{0.08\linewidth}p{0.08\linewidth}p{0.08\linewidth}p{0.08\linewidth}p{0.08\linewidth}p{0.08\linewidth}p{0.08\linewidth}p{0.085\linewidth}p{0.08\linewidth}}
\hline
&  &   $\Gamma=\Gamma_0$&&&$\Gamma=\Gamma_0 T$&&&$\Gamma=\Gamma_0 T^3$&\\
\hline
& $P_1$ &  $P_2$ &$P_3$&$P_1$ &  $P_2$ &$P_3$&$P_1$ &  $P_2$ &$P_3$ \\
\hline
m& 50.0 &  60.0 &70.0&50.0&60.0&70.0&100&100&100 \\
$\Gamma_0(\times 10^{-6})$& 0.010 &  0.010 &0.010&200&200&200&$50.0\times10^9$&$60.0\times10^9$&$70.0\times10^9$ \\
$n_s(\times 10^{-2})$& 96.3 &  96.7 &97.1&97.6&96.7&97.1&97.7&97.3&96.5 \\
$r(\times 10^{-2})$& 1.37 & 3.98 &9.12&0.510&1.43&3.20&2.40&2.00&1.71 \\
$Q(\times 10^{-4})$& 7.99 &  1.61 &0.458&20.9&6.19&2.38&32.1&66.6&123 \\
$H(\times 10^{-6})$& 4.16 & 20.7 &72.8&0.930&5.26&20.5&72.1&72.1&72.1 \\
$T(\times 10^{-5})$& 4.85 &  6.92 &9.12&2.92&4.89&7.31&24.0&28.8&33.7 \\
$T/H$& 11.6 &  3.34 &1.25&31.3&9.29&3.56&3.33&4.00&4.67 \\
\hline
\end{tabular}
\end{table*}

Few important points are in order here. First, unlike the cold scenario, the exponential potential of eq. (\ref{eq:VintermsOfphi_weak}) will not remain always the one on which the inflaton moves, because it was obtained assuming the weak limit ($Q \ll 1$). The issue here is that $Q$ is a dynamical quantity which, thus, evolves in time, as can be seen in the Figure-table (\ref{table:weak_formuals}). Starting from a value where $Q \ll 1$, the inflaton field tends to grow going down the potential surface, but then $Q$ evolves quickly to be exponentially larger (smaller) in the first two cases (third case) of constant/$T$-linearly dependent ($T$-cubically dependent) dissipation factor, and consequently only in the third case the exponential shape remain valid for all the inflaton values.

To fix the ideas, we took the parameters stated in the Figure-table (\ref{TableQVweak_no_exit}) for the three cases of the dissipative factor $\Gamma$, and  illustrated the corresponding forms of $Q$ and $V$, in which place the green dot denotes the horizon crossing, where the observables are evaluated and found to meet the observational constraints, whereas the red point designates the bound of the weak limit region, corresponding to ($Q=1$), signalling where the validity of the weak dissipation assumption ceases to be maintained. Thus, the portion of the exponential graph describing the weak limit is bounded by the red point, and one should check for consistency, upon introducing an exit scenario as will be done in the next section, in that the end of inflation, assuming a weak limit regime, should remain on this portion not reaching the corresponding red point.
\begin{table*}
\begin{tabular}{cccc}
\toprule
$\Gamma =$ & $\Gamma_0$ & $\Gamma_0 T$ & $\Gamma_0 T^3$  \\
\midrule
Parameters & $m=45, \alpha = 75, \Gamma_0=10^{-8}$ & $m=42, \alpha = 75, \Gamma_0=10^{-2}$ & $m=100, \alpha = 75, \Gamma_0=5 \times 10^{4}$  \\
\midrule
Q & \includegraphics[width=5cm]{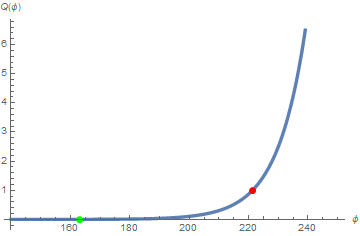} &
\includegraphics[width=5cm]{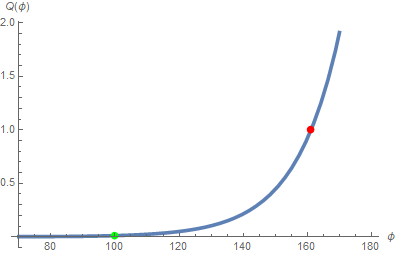}&
\includegraphics[width=5cm]{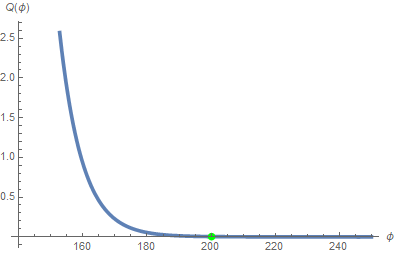} \\
\midrule
V & \includegraphics[width=5cm]{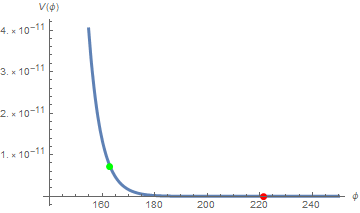} &
\includegraphics[width=5cm]{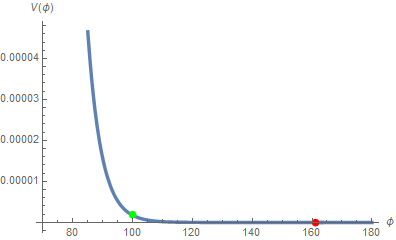}&
\includegraphics[width=5cm]{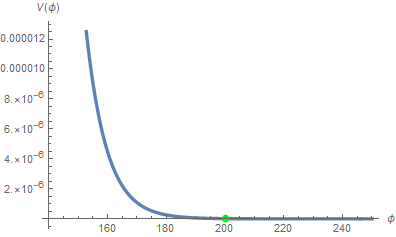} \\
\midrule
Green $(\Phi_*, Q, V)$ &
 $(163, 2 \times 10^{-3}, 7 \times 10^{-12})$ &
$(99.8, 11 \times 10^{-3}, 1.8 \times 10^{-6})$ &
$(200, 3 \times 10^{-3}, 1.6 \times 10^{-8})$ \\
\midrule
Observables ($n_s, r$) & ($0.976, 0.007$) & ($0.958,0.109$) & ($0.977,0.024$)  \\
\midrule
Red $(\Phi_*, Q, V)$ &
 $(221, 1, 3 \times 10^{-17})$ &
  $(161, 1, 2.8 \times 10^{-12})$ &
$\nexists$ \\
\bottomrule
\end{tabular}
\caption{Warm inflation parameters in the weak limit regime ($Q\ll 1$) leading to acceptable observables, in the three cases of dissipative factor $\Gamma$. `Green' corresponds to horizon crossing, whereas 'Red' denotes where the weak limit regime stops ($Q =1$).\label{TableQVweak_no_exit}}
\end{table*}

Second, the choice of benchmarks is not exhaustive. Actually, and as will be shown in next section, the parameter space accommodating the observational constraints is not too much limited, and no need to resort to fine tuning in order to meet the constraints on $(n_s,r)$. One can start analytically by expanding in powers of $\phi_*$ and see that generically $\Gamma_0$ appears raised to some positive power in the denominator of the leading order of $r$, which means that one needs large values of $\Gamma_0$, in order to meet $(r \ll 1)$, but then these large values of $\Gamma_0$ can not satisfy the condition on $n_s$ (look at the expressions of $r, n_s$ in Table \ref{table:weak_formuals}) with $\phi_*$ small. So, one searches for large values of $\phi_*$ and once we find an acceptable value we can refine the search around it, while changing accordingly the other parameters. Upon carrying out the scan, one finds a non-negligible region in the parameters space accommodating the phenomenological constraints.

Third, the issue of $Q$ being dynamical will not change the conclusion that the inflation in this model does not have an end, because ($\epsilon, \eta, \beta$) are constant and the conditions (Eq. \ref{slowrollconditions}) remain satisfied if they are met at horizon crossing.

\subsection{Power-Law inflation in the Strong Dissipation Limit}
Similarly, in the strong dissipation limit, we can identify  the potential responsible for the power-law expansion as we did in the weak dissipation limit. However, it is interesting to note that, unlike the weak case where the potential was exponential, the strong dissipation limit does not have a single-form potential; instead, we can find different forms of potential for the different types of dissipation parameter $\Gamma(T)$.

Starting with the case $\Gamma=\Gamma_0$, and  substituting Eq.(\ref{eq:Vintermsoft}) into the integration of Eq.(\ref{eq:dotphi_SRA}), we find
\begin{equation}
\label{eq:phi}
\phi= -\frac{2 \sqrt{6} m}{\sqrt{{\Gamma_0}} \sqrt{t}}
\end{equation}
As a result, the potential $V$ is given in terms of the field $\phi$ as
\begin{eqnarray}
\label{eq:VintermsOfphi_strong_Gamma0}
V=\frac{{\Gamma_0}^2 }{2^6\times 3 m^2} {\phi}^4 &,& \mbox{for } \Gamma = \Gamma_0 = constant.
\end{eqnarray}
The potential (\ref{eq:VintermsOfphi_strong_Gamma0})appears as a quartic one in the field.

As to the other cases of $\Gamma(T)$, we can find different potentials using the same strategy:
\begin{eqnarray}
V(\phi)=V_0{\phi}^8: V_0=\frac{{\Gamma_0}^4 }{2^{21}\times 3^2 {\alpha} m^5} &,& \mbox{for } \Gamma = \Gamma_0 T \label{strong_gamma0T_potential}
\\
V(\phi)= V_0 {\phi}^{-8}: V_0=\frac{2^{23} \times 3^2 {\alpha}^3 m^7}{{\Gamma_0}^4 } &,& \mbox{for } \Gamma = \Gamma_0 T^3 \label{strong_gamma0T3_potential}
\end{eqnarray}
The potential of Eq. \ref{strong_gamma0T_potential} (\ref{strong_gamma0T3_potential}) is just  a monomial with positive (negative) power.

Table (\ref{tab:strong_formulas}) shows the slow-roll parameters in each case, along with the dissipation effectiveness coefficient $Q$, in addition to the observables $n_s$ and $r$. From the table, we can see that the inflation has no end under the slow-roll approximation because the slow-roll parameters $\varepsilon$, $\eta$, and $\beta$ have all the same dependence ($\propto \phi^{-2}$) on the field, as  that of the dissipation effectiveness coefficient $Q$ so that their ratio becomes field-independent.

The table shows that for the case $\Gamma=\Gamma_0$, where the potential $V \propto \phi^4$, the spectral index equals zero, which is excluded by experimental surveys in which $n_s < 1$. On the other hand, for $\Gamma=\Gamma_0 T$, we have $n_s >1$, which is also excluded. Finally, looking at the case  ($\Gamma=\Gamma_0 T^3$), we can see that $n_s<1$ and the model can successfully accommodate the experimental results.

\begin{figure*}
  \includegraphics[width=1\textwidth]{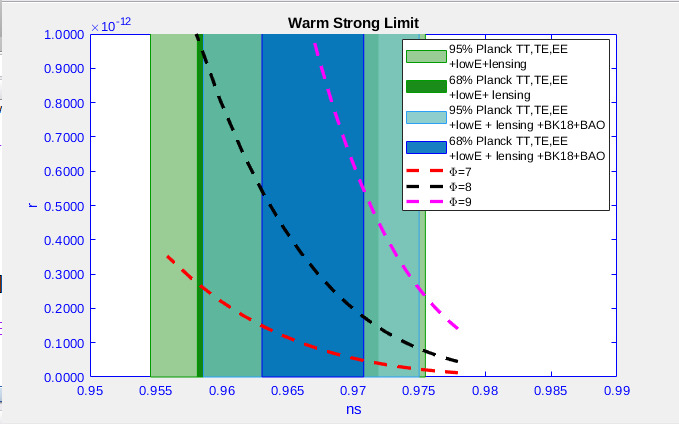}
\caption{A plot of $r$ and $n_s$ under the strong dissipation limit in the case of $\Gamma=\Gamma_0 T^3$. All lines correspond to $\Gamma_0=3\times10^{7}$, while $\phi_\ast = 7,8,9$ for the red, green, and blue lines, respectively. The parameter $m$ is scanned in the range [100-200] in all cases.}
\label{fig:ns_r_strong}       
\end{figure*}

Figure (\ref{fig:ns_r_strong}) shows the theoretical results of the model under the strong dissipation limit for the case ($\Gamma=\Gamma_0 T^3$). As we see, the results are in perfect agreement with the empirical ones. The parameter $m$ is scanned between 100 and 200 for specific values of $\phi_*$ for each line; most results are within the $68\%$ confidence level.

Table (\ref{tab:values_strong}) shows some benchmark points in the strong limit with ($\Gamma=\Gamma_0 T^3$). In this table, we stated the values of the observables, $n_s$ and $r$, in addition to those of  $Q$, where we should keep in mind that the case is performed under the strong dissipation limit $Q\gg1$. The table also shows both $H$ and $T$ in which $T>H$ which is usually required in the warm inflation.

\begin{table*}
\caption{Main specific characteristics of the warm model in the strong dissipative regime for different cases of dissipation rate $\Gamma(T)$.}
\label{tab:strong_formulas}       
\begin{tabular}{p{0.2\linewidth}p{0.25\linewidth}p{0.25\linewidth}p{0.2\linewidth}}
\hline\noalign{\smallskip}
 & $\Gamma=\Gamma_0$ &  $\Gamma=\Gamma_0 T$ & $\Gamma=\Gamma_0 T^3$  \\
\noalign{\smallskip}\hline\noalign{\smallskip}
$\varepsilon$ & $\frac{8}{{\phi}^2}$ &  $\frac{32}{{\phi}^2}$ & $\frac{32}{{\phi}^2}$\\
$\eta$ & $\frac{12}{{\phi}^2}$ &  $\frac{56}{{\phi}^2}$ & $\frac{72}{{\phi}^2}$ \\
$\beta$ & $0$ & $\frac{16}{{\phi}^2}$ & $\frac{48}{{\phi}^2}$ \\
$Q$ & $\frac{8m}{{\phi}^2}$ &  $\frac{32m}{{\phi}^2}$ & $\frac{32 m}{\phi^2}$ \\
$T$ &$\frac{\sqrt{{\Gamma_0}} {\phi}}{2\ 2^{3/4} \sqrt[4]{3} \sqrt[4]{{\alpha}} m^{3/4}}$  &$\frac{\Gamma_0 \phi^2}{32 \sqrt{6}\sqrt{\alpha}m^{3/2}}$  &$\frac{32 \sqrt{6} \sqrt{{\alpha}} m^{3/2}}{{\Gamma_0} {\phi}^2}$\\
$H$ &$\frac{\Gamma_0 \phi^2}{24 m}$ & $\frac{\Gamma_0^2 \phi^4}{3072 \sqrt{6}\sqrt{\alpha}m^{5/2}}$ &   $\frac{2048 \sqrt{6} {\alpha}^{3/2} m^{7/2}}{{\Gamma_0}^2 {\phi}^4}$\\
$n_s-1$ & $0$ &  $\frac{0.4}{m}$ & $-\frac{4.415}{m}$ \\
$r$ & $\frac{\sqrt[4]{{\alpha}} \sqrt{{\Gamma_0}} {\phi}^4}{6\ 2^{3/4} \sqrt[4]{3} \sqrt{\pi } m^{11/4}}$ &  $\frac{5.3\times 10^{-6} \Gamma_0 \phi^{9.63}}{m^{5.815}}$ &  $\frac{{4.4\times 10^{-6} {\alpha} {\phi}^{9.66}}}{{\Gamma_0} m^{4.83}}$ \\

\noalign{\smallskip}\hline
\end{tabular}
\end{table*}

\begin{table}
\begin{tabular}  {p{0.21\linewidth}p{0.21\linewidth}p{0.21\linewidth}p{0.21\linewidth}}
\hline\noalign{\smallskip}
 & $P1$ &  $P2$ & $P3$  \\
\noalign{\smallskip}\hline\noalign{\smallskip}
m & $100$ &  $150$ & $200$\\
$n_s$ & $0.956$ &  $0.970$ & $0.977$\\
$r(\times 10^{-13})$ & $3.51$ &  $0.495$ & $0.123$ \\
$Q$ & $65.3$ & $97.9$ & $131$ \\
$T(\times10^{-4})$ & $4.61$ &  $8.48$ & $13.0$ \\
$H(\times10^{-5})$ &$1.50$  &$6.23$  &$17.0$\\
$T/H$ &$30.6$ & $13.6$ &   $7.65$\\
\noalign{\smallskip}\hline
\caption{Benchmark points in the case of strong dissipation limit in the case of $\Gamma=\Gamma_0 T^3$ with  $\Gamma_0 = 3\times 10^7$ and $\phi_*=7$.}
\label{tab:values_strong}       
\end{tabular}
\end{table}

Again, one should determine the validity range of the strong limit regime. We restrict this analysis to the third case of cubically $T$-dependent dissipation factor, as the other two cases are rejected phenomenologically, and the Figure-table (\ref{TableFigureQVstrong_no_exit}) shows the plots of $Q, V$ vs $\phi$ for a benchmark accommodating the observational constraints. The two points (green, red) designate respectively the horizon crossing and strong limit regime boundary, and so the portion of the $V$-curve to the left of the red point (the dashed part) is the part representing the strong limit regime potential.

\begin{table}
\begin{tabular}{cc}
\toprule
$\Gamma =$ &  $\Gamma_0 T^3$  \\
\midrule
Parameters & $m=100, \alpha = 75, \Gamma_0=3 \times 10^{7}$  \\
\midrule
Q &
\includegraphics[width=5cm]{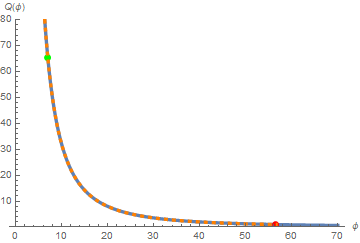} \\
\midrule
V & \includegraphics[width=5cm]{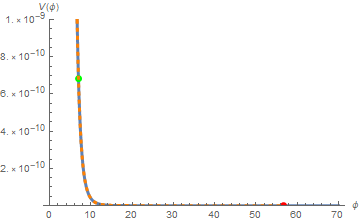} \\
\midrule
Green $(\Phi_*, Q, V)$ &
$(7, 65.3, 7 \times 10^{-7})$ \\
\midrule
Observables ($n_s, r$) & ($0.956,2 \times 10^{-10}$)   \\
\midrule
Red $(\Phi_*, Q, V)$ &
 $(56, 1, 4 \times 10^{-17})$ \\
\bottomrule
\end{tabular}
\caption{Warm inflation parameters in the strong limit regime ($Q\gg 1$) leading to acceptable observables, in the case of cubically $T$-dependent dissipative factor $\Gamma= \Gamma_0 T^3$. `Green' corresponds to horizon crossing, whereas `Red' denotes where the strong  limit regime stops ($Q =1$).  \label{TableFigureQVstrong_no_exit}}
\end{table}

\section{Possible graceful  exit }
\label{sec:Possible graceful  exit}

The purpose of this section is to suggest a model modification, representing a small perturbation, that would enable the inflationary phase to be exited. Following the dynamically motivated approach, our strategy will be the same as the one adopted in the power-law model of inflation, where we started from the Hubble parameter and ended in determining the potential. However, we assume now a small perturbation in the Hubble parameter as,
\begin{equation}\label{eq:Hwithexit}
H=\frac{m}{t}+\zeta
\end{equation}
where $\zeta$ acts as an affine perturbation, typically, as we shall see, of the order of $10^{-7}$ with negative sign.  According to this modification, the power-law expansion (Eq. \ref{eq:PowerLaw_H}) appears as the first term of Eq. \ref{eq:Hwithexit}, inversely proportional to time, which dominates during the early stage of the universe over the affine perturbation. Specifically, we restrict our study here to the weak dissipation limit, as we found that the perturbed potential was well determined analytically in this case, unlike the strong case where it needed to be determined numerically.  Since the unperturbed potential in this limit is exponential in the scalar field, similar to the cold paradigm, we expect, and will show it to be the case, that finding an exit solution by perturbing will offer a similar way out for the problems of the cold power-law inflation.

A similar treatment to that of section [\ref{sect:PoweWeak}] reveals that the potential, in the presence of the affine perturbation $\zeta$, is modified to:
\begin{equation}\label{eq:U}
U=\sum_{j=0}^{j=2} A_j e^{-j \frac{\phi}{\sqrt{2m}}}
\end{equation}
where:
\begin{equation}
A_0= 3 \zeta^2 ,\qquad A_1=6 m \zeta ,\qquad A_3= 3 m^2
\end{equation}
Note that, in the large field limit, we have a constant positive term ($A_0 \propto \zeta^2$) dominant corresponding to a Minkowski universe with a tiny cosmological constant.

\begin{figure}
  \includegraphics[width=0.5\textwidth]{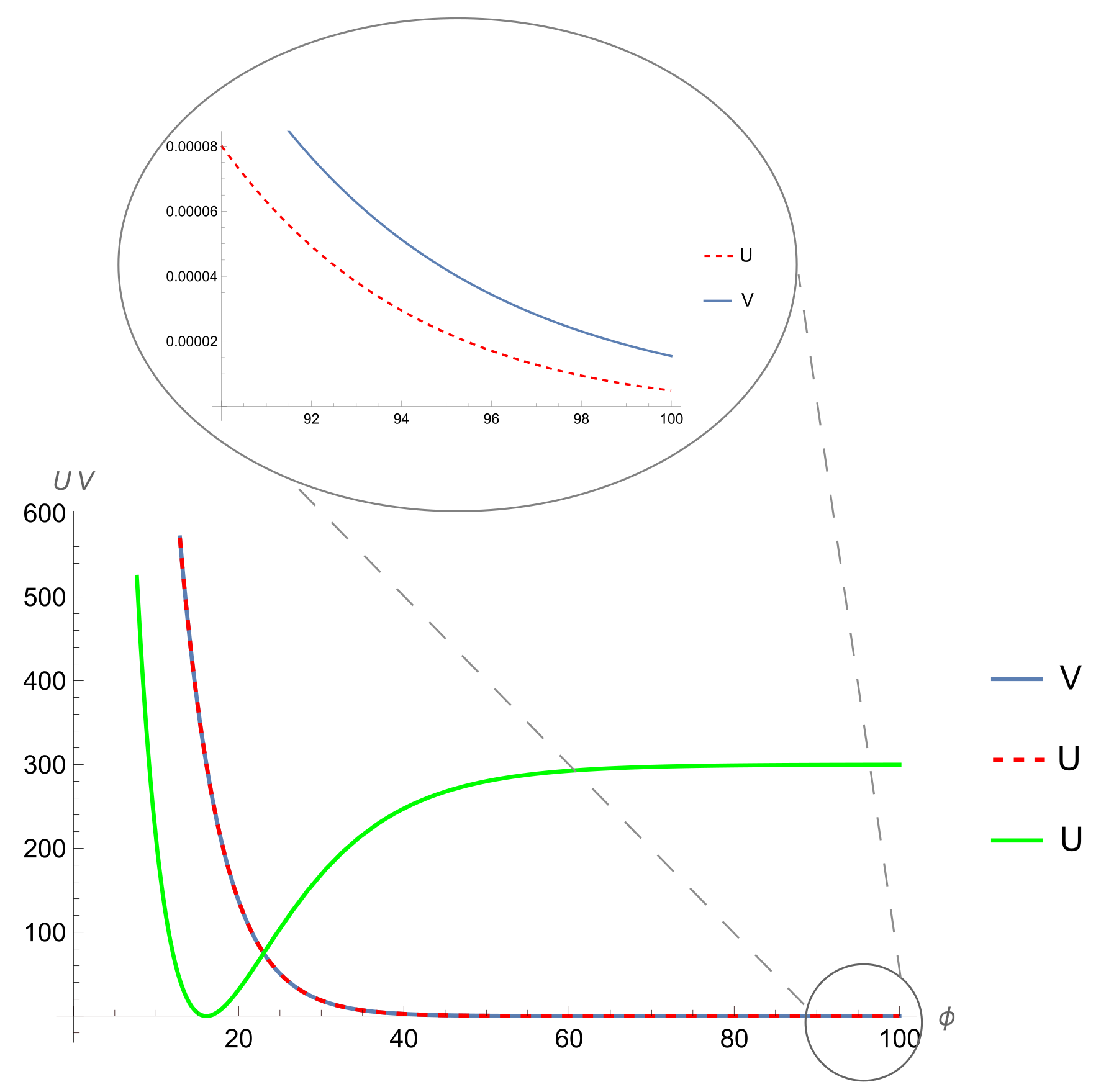}
\caption{Unperturbed (Perturbed) Potential $V$ ($U$) as function of the scalar field $\phi$, with $m=50$. Blue and dashed red lines correspond respectively to the potential $V$ and $U$ with $\zeta=-0.001$. Zoomed area shows the tiny deviations between the two potentials in the range $\phi \in [90-100]$. The green line represents the potential $U$ for $\zeta=-10$, where we see a tangible change in the potential form whenever $|\zeta| >1$ . }
\label{fig:potential_weak_exit}       
\end{figure}

As shown in Figure (\ref{fig:potential_weak_exit}), the potentials (\ref{eq:U}) and (\ref{eq:VintermsOfphi_weak}) are the same for $|\zeta| \ll 1$ but differ palpably for $|\zeta| > 1$, which is not considered in this study. However, we see in the figure's zoomed area that the potentials are not completely identical in the case of $\zeta \ll 1$. Although the difference is slight, it is, however, sufficient to achieve the necessary requirements in order to end the inflationary epoch. Actually, and as opposed to pure exponential potential, the slow roll parameters appear now as functions of the scalar field,
\begin{equation} \label{varepsilon_withExit}
\varepsilon=\frac{m}{\left({\zeta} e^{\frac{{\phi}}{\sqrt{2} \sqrt{m}}}+m\right)^2}
\end{equation}
\begin{equation} \label{eta_withExit}
\eta=\frac{{\zeta} e^{\frac{{\phi}}{\sqrt{2} \sqrt{m}}}+2 m}{\left({\zeta} e^{\frac{{\phi}}{\sqrt{2} \sqrt{m}}}+m\right)^2}
\end{equation}
\begin{equation} \label{beta_withExit}
\beta=C \frac{ \left(2 {\zeta} e^{\frac{{\phi}}{\sqrt{2} \sqrt{m}}}+m\right)}{\left({\zeta} e^{\frac{{\phi}}{\sqrt{2} \sqrt{m}}}+m\right)^2}
\end{equation}
with $C= 0$ for $\Gamma=\Gamma_0$  , $C= 1/3$ for $\Gamma=\Gamma_0 T$ ,and  $C= 3$ for $\Gamma=\Gamma_0 T^3$. We see from Eqs. (\ref{varepsilon_withExit}-\ref{beta_withExit}) that $\zeta$ should be of negative sign, in order to allow for an exit scenario, otherwise ($\varepsilon, \eta, \beta$) keep decreasing when the inflaton rolls down the potential surface.

Figure (\ref{fig:ns_r_weak_warm}) depicts the observables $n_s$ and $r$, which are represented by dashed lines in this perturbed case. We fixed all parameters to those of the pure unperturbed exponential potential case, except with $\zeta=-1\times10^{-7}$ for both $\Gamma=\Gamma_0$ and $\Gamma=\Gamma_0 T$, and $\zeta=-1.5\times10^{-6}$ for $\Gamma=\Gamma_0 T^3$.

Despite the slight deviations, the results of ($n_s, r$) are pretty similar in the perturbed and unperturbed cases. In contrast, unlike the unperturbed case, the perturbed results now correspond to $N\in[81-392]$ folds for $\Gamma=\Gamma_0$ and $N\in[66-302]$ folds for $\Gamma=\Gamma_0 T$, whereas $N$ is still not applicable for $\Gamma=\Gamma_0 T^3$ since, as we shall see, there is no end to inflation in the weak regime. The effect of the parameter $\zeta$ is shown in table (\ref{table:weak_with_exit}), where we take specific points with the same input parameters of ($m, \Gamma_0$ and $\Phi_*$) but with different values of $\zeta$. We present the corresponding number of e-folding which is less than the acceptable upper bound of 80 \cite{liddle}. As expected, as $\zeta$ decreases in magnitude, there is a decrease in the difference between the pure ``unperturbed'' exponential potential and the modified ``perturbed" exponential potential and vice versa. However, we note that in the third case of cubically $T$-dependent dissipation factor there is no end scenario and the corresponding efoldings number is not applicable.

\begin{table*}[ht]
\caption{Main specific characteristics of the ``perturbed" warm model in the weak dissipative regime with modified exponential potential.}\label{table:weak_with_exit}
\centering
\begin{tabular}{p{0.03\linewidth}p{0.05\linewidth}p{0.05\linewidth}p{0.05\linewidth}p{0.05\linewidth}p{0.05\linewidth}p{0.05\linewidth}p{0.08\linewidth}p{0.05\linewidth}p{0.05\linewidth}p{0.05\linewidth}p{0.085\linewidth}p{0.05\linewidth}}
\hline
&  &   &$\Gamma=\Gamma_0$&&&$\Gamma=\Gamma_0 T$&&&&$\Gamma=\Gamma_0 T^3$\\
\hline
&$\zeta(10^{-9})$& $n_s$ &  $r$ &$N$&$\zeta(10^{-9})$&$n_s$ &  $r$ &$N$&$\zeta(10^{-6})$&$n_s$ &  $r$ &$N$ \\
\hline
$V$&0&0.976& 0.0070 &  N/A &0&0.976&0.0051&N/A&0&0.977&0.024&N/A \\
&&&  &   &&&&&&\\
&-167&0.971& 0.0079 &  60 &-145.6&0.977&0.0057&50&-21.6&0.949&0.024&N/A \\
$U$&-129.8&0.972& 0.0077 &  70 &-115.4&0.976&0.0055&60&-18.8&0.951&0.024&N/A \\
&-103.5&0.973& 0.0075 &  80 &-92.2&0.975&0.0054&70&-16.5&0.954&0.024&N/A \\
\hline
\end{tabular}
\end{table*}

Actually, and as was done in the unperturbed scenario, one should check for consistency that the weak limit regime ($Q \ll 1$) is well respected for the dynamical variable $Q$ through the whole of inflation. For this, we show in the Figure-table (\ref{TableQVweak_with_exit})
the dissipative coefficient $Q$ and the potential $U$ for the three cases of the dissipative factor $\Gamma$, for some benchmark points in the parameter space meeting the observational constraints of ($n_s, r$) and of $A_s$ \footnote{The curvature perturbation spectrum has been measured by PLANCK (WMAP) at
$68\%$ Confidence Level at the fixed wave number $k_{\star} = 0.05(0.002) \mbox{ Mpc}^{-1}$ as
\cite{planck2013}(\cite{wmap2013})
\begin{eqnarray}
\label{A_s_Constraints}
A_s \in [2.136,2.247] \times 10^{-9} &,& \left( \in [2.349,2.541] \times 10^{-9}\right)
\end{eqnarray}
and thus the model seeks at least to reproduce a comparable order of magnitude ($ \lessapprox 10^{-9}$).} at the horizon crossing, where, as in the unperturbed scenario, the green and red dots designate respectively the horizon crossing, where the observables are evaluated, and the stoppage of the weak limit regime, i.e. the ($Q=1$)-boundary. However, a yellow dot is added here to denote now the end of inflation where one of the conditions of Eq. (\ref{slowrollconditions}) breaks down (whichever first). In all of these cases, the inflaton field would roll to the right, but in the  third case of cubically $T$-dependent dissipation factor there is no end of inflation consistent within the weak regime, since the slow roll conditions of Eq. (\ref{slowrollconditions}) are maintained due to ($1+Q$) being far larger than ($\varepsilon, \eta, \beta$). The consistency check amounts to ($Q \ll 1$) at the yellow dot, which, as the figure table (\ref{TableQVweak_with_exit}) shows, is met indeed. The red dot determines the portion (dashed curve) of the potential curve which represents correctly the inflaton potential.

\begin{table*}
\hspace{-1.7cm}
\begin{tabular}{cccc}
\toprule
$\Gamma =$ & $\Gamma_0$ & $\Gamma_0 T$ & $\Gamma_0 T^3$  \\
\midrule
Parameters & $m=45,  \Gamma_0=10^{-8}, \zeta = -1.7 \times 10^{-7}$ & $m=50,  \Gamma_0=2 \times 10^{-4}, \zeta = -1.5 \times 10^{-7}$ & $m=100,  \Gamma_0=79 \times 10^{3}, \zeta = -1.5 \times 10^{-5}$  \\
\midrule
Q & \includegraphics[width=5cm]{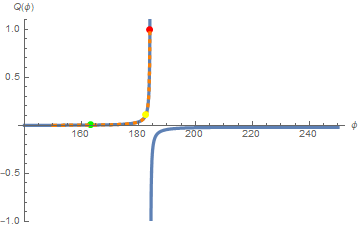} &
\includegraphics[width=5cm]{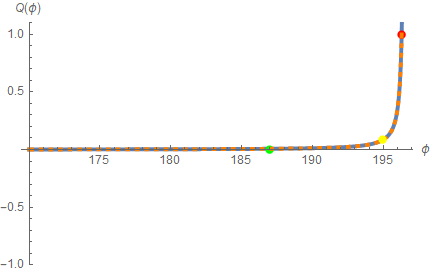}&
\includegraphics[width=5cm]{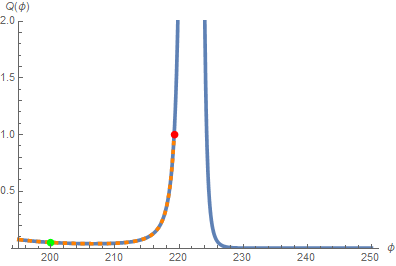} \\
\midrule
U & \includegraphics[width=5cm]{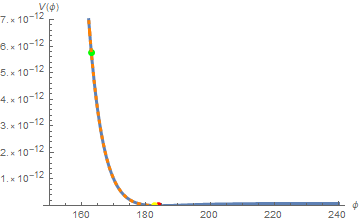} &
\includegraphics[width=5cm]{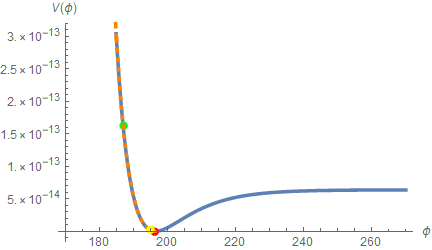}&
\includegraphics[width=5cm]{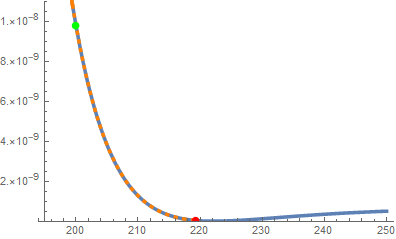} \\
\midrule
slow roll parameters &
\includegraphics[width=5cm]{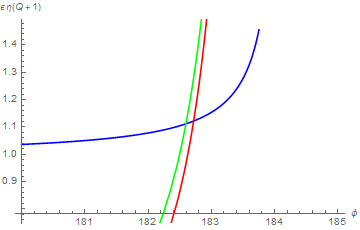} &
\includegraphics[width=5cm]{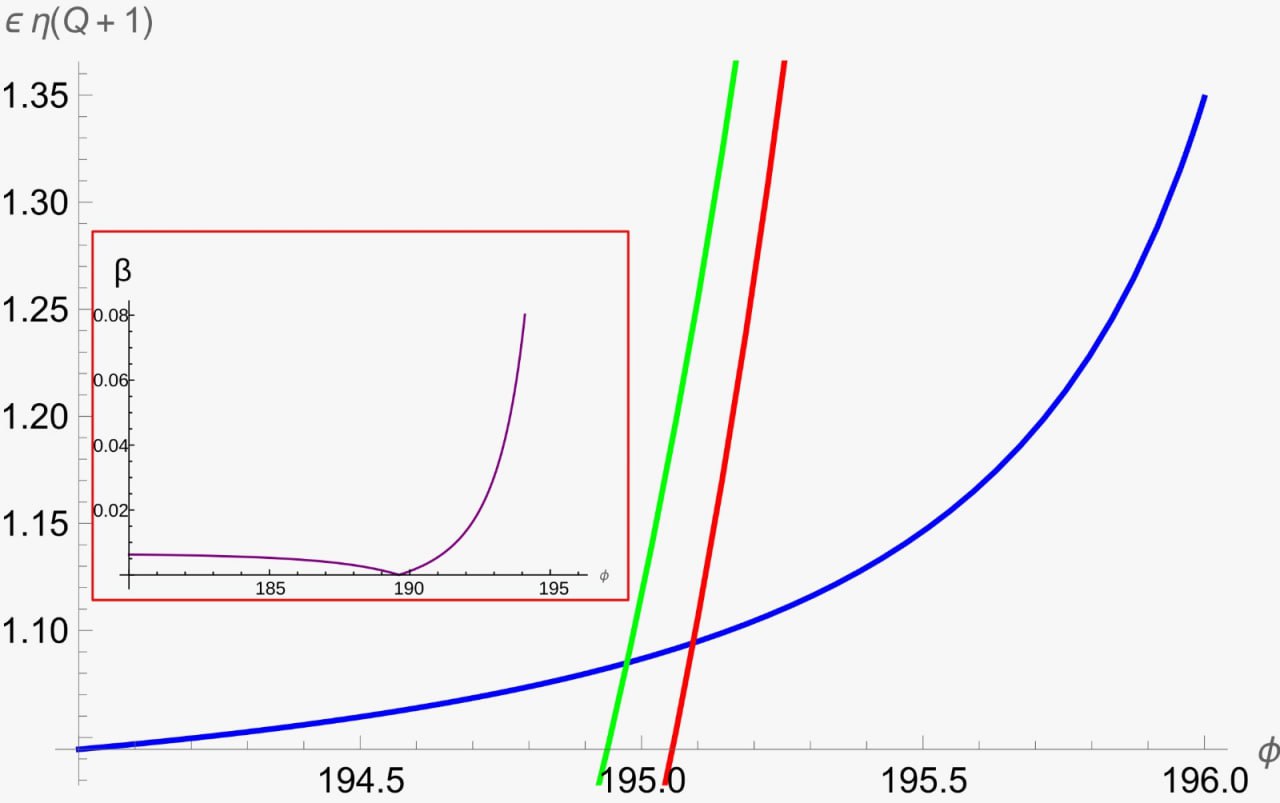} &
\includegraphics[width=5cm]{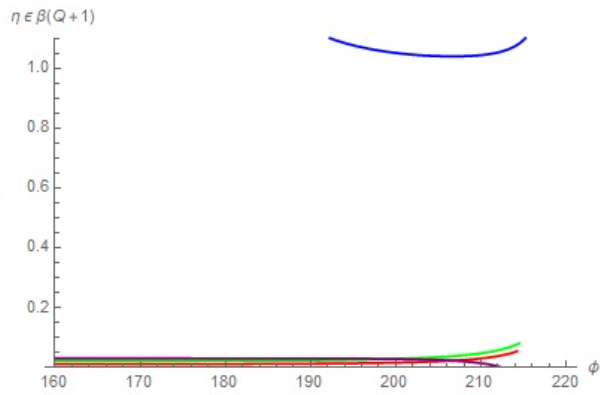} \\
\midrule
Green $(\Phi_*, Q, U)$ &
 $(163, 0.002, 5.7 \times 10^{-12})$ &
$(187, 0.007, 1.6 \times 10^{-13})$ &
$(200, 0.05, 9.7 \times 10^{-9})$ \\
\midrule
Observables ($n_s, r, A_s$) & ($0.972, 0.008, 5\times 10^{-11}$) & ($0.977, 0.006, 2\times 10^{-11}$) &($0.914, 0.015, 5\times 10^{-8}$)  \\
\midrule
Yellow $(\Phi_*, Q, U)$ &
 $(182, 0.11, 2.7 \times 10^{-15})$ &
  $(194, 0.08, 1.8 \times 10^{-15})$ &
$\nexists$ \\
\midrule
efoldings $N$ & $60$ & $50$ & $\nexists$ \\
\midrule
Red $(\Phi_*, Q, U)$ &
 $(184, 1, 3 \times 10^{-17})$ &
  $(196, 1, 4 \times 10^{-17})$ &
$(219, 1, 3.7 \times 10^{-11})$ \\
\bottomrule
\end{tabular}
\caption{Warm inflation parameters, within the perturbative scenario with $\alpha = 75$, in the weak limit regime ($Q\ll 1$) leading to acceptable observables, in the three cases of dissipative factor $\Gamma$. `Green' dot corresponds to horizon crossing, `Yellow' dot corresponds to end of inflation whereas `Red' dot denotes where the weak limit regime stops ($Q =1$). For the slow roll parameters, `Blue' curve corresponds to ($1+Q$), `Red' curve corresponds to $\varepsilon$, `Green' curve corresponds to $\eta$ whereas `Purple' curve corresponds to $\beta$. \label{TableQVweak_with_exit}}
\end{table*}

To conclude on the graceful exit in warm inflation \cite{2005.01122}, we present also in the Figure-table (\ref{TableQVweak_with_exit}) the plots of ($1+Q$) in blue, $\varepsilon$ in red, $\eta$ in green and $\beta$ in purple for the considered benchmark points in three dissipation cases. We see that $\phi_{e}$, the inflaton value at end of inflation, in the first case of constant $\Gamma$ occurs at ($\phi_e \sim 182$) and is determined by the condition ($1+Q = \eta$), which helps also in the second case of linearly $T$-depenedent $\Gamma$, where $\beta$ is too small, to determine that ($\phi_e \sim 194$). In the third case of cubically $T$-dependent $\Gamma$, the curves of ($\varepsilon, \eta, \beta$) are with too small values, that do not intersect those of $(1+Q)$, so no consistent end of inflation within the weak limit region in this case.

Again, as in the unperturbed case, the choice of $\phi_*$ does not need to be fine tuned, but a simple scan for large values of $\phi_*$ and suitable choices of the other parameters can hit an accommodating data solution, in the neighbourhood of which, changing appropriately the other parameters, one can find many other solutions. In Table (\ref{WarmWeakExitFirstCase_benchmarks}) we list several benchmark points with the corresponding observables of ($n_s, r, A_s$), and constraints on ($Q \ll 1$, $T/H \gg 1$ and $N \geq 50$) for the constant case ($\Gamma =\Gamma_0$). For the second case ($\Gamma= \Gamma_0 T$), we checked also the existence of considerable acceptable region in the parameter space, but we did not study the third case ($\Gamma = \Gamma_0 T^3$) as there was no end to inflation.

\begin{table*}[ht]
\caption{Benchmark points in the case of weak dissipation limit for the case of constant dissipation factor $\Gamma=\Gamma_0$, with $\alpha=75$, and $Q_e$ denoting $Q$ at end of inflation.   }
\label{WarmWeakExitFirstCase_benchmarks}
\centering
\begin{tabular}{c|cccc|ccc|ccc}
\hline
&\multicolumn{4}{c} {Parameters}  &   \multicolumn{3}{c} {Observables}& \multicolumn{3}{c} {Constraints} \\
\hline
& $m$ & $\Gamma_0$ & $\phi_*$ & $\zeta$ &
 $n_s$ & $r$ & $A_s$ &
  $N$ & $Q_{e}$ & $T/H$ \\
\hline
$1$ &
$80$ & $10^{-9}$ & $200$ & $-1.035 \times 10^{-7}$ &
$0.974$ & $0.0355$ & $1.04 \times 10^{-10}$ &
$292$ & $0.035$ & 2.87\\
$2$ &
$60$ & $10^{-9}$ & $200$ & $-1.035 \times 10^{-7}$ &
$0.954$ & $0.006$ & $1.16 \times 10^{-11}$ &
$63.5$ & $0.021$ & 28.8\\
$3$ &
$143$ & $10^{-8}$ & $300$ & $-1 \times 10^{-6}$ &
$0.966$ & $0.006$ & $1.05 \times 10^{-10}$ &
$55.8$ & $0.035$ & 20.7\\
$4$ &
$150$ & $10^{-8}$ & $300$ & $-1 \times 10^{-6}$ &
$0.975$ & $0.008$ & $3.24 \times 10^{-10}$ &
$108$ & $0.036$ & 11.4\\
$5$ &
$241$ & $10^{-9}$ & $400$ & $-1 \times 10^{-6}$ &
$0.974$ & $0.008$ & $9.73 \times 10^{-11}$ &
$100$ & $0.005$ & 9.6\\
$6$ &
$234$ & $10^{-9}$ & $400$ & $-1 \times 10^{-6}$ &
$0.958$ & $0.007$ & $3.7 \times 10^{-11}$ &
$55$ & $0.004$ & 15.5\\
\hline
\end{tabular}
\end{table*}

    A further interesting fact is that the potential of Eq. (\ref{eq:U}), which is responsible for the expansion described by Eq. (\ref{eq:Hwithexit}), may have contributed to a transition into another cosmological phase after the inflationary stage. Let us assume Eq. (\ref{eq:Hwithexit}) to be valid even after the inflationary era, then using $\ddot{a} = a (\dot{H} + H^2)$, one can determine the signs of $\dot{a}, \ddot{a}$ in terms of the physical comoving time $t$, as in table (\ref{sign}), where we do not extrapolate beyond $t_2 = \frac{m}{|\zeta|}$ as this corresponds to a contracting universe unlike ours.

    \begin{table}[ht]
\centering
\begin{tabular}{c|ccccc}
\hline
$t$ & $0$ && $t_1 = \frac{m-\sqrt{m}}{|\zeta|}$ & & $t_2 = \frac{m}{|\zeta|}$ \\
\hline
$\dot{a}$ && $+$ && $+$ & $0$ \\
$\ddot{a}$ && $+$ &$0$& $-$ &  \\
\hline
\end{tabular}
\caption{Signs of $\dot{a}$ and $\ddot{a}$.}\label{sign}
\end{table}
Thus, one can justify moving at $t_1 = \frac{m-\sqrt{m}}{|\zeta|}$ from the inflationary stage ($\ddot{a} >0$) to a decelerated phase ($\ddot{a} <0$) which would correspond to matter dominated era. However, the decelerating phase here corresponds to an e-folding number increase by just around $0.5$ \footnote{This comes because $\Delta (N_e) = \int_{t_1}^{t_2} H dt = \int_{t_1}^{t_2} (\frac{m}{t} + \zeta) dt = 0.5 + \mathcal{O}(1/\sqrt{m})$.} which can not account for the whole matter dominated phase. The figure table (\ref{Transient}) shows $H$ and $\ddot{a}$ versus the e-foldings number $N$, where the red $X$ mark (sky blue dot) correspond to $t_2$ ($t_1$), for the  choice ($m=45, \zeta=-1.035 \times 10^{-7}$) leading to:
\begin{eqnarray}
79.86 \approx N= & \leftrightarrow H=0 \leftrightarrow& t_2=\frac{m}{|\zeta|}=4.35 \times 10^8 \mbox{ s} \nonumber \\
79.36 \approx N= & \leftrightarrow \ddot{a}=0 \leftrightarrow& t_1=\frac{m-\sqrt{m}}{|\zeta|}=3.70 \times 10^8 \mbox{ s} \nonumber
\end{eqnarray}

\begin{table}
\begin{tabular}{cc}
\toprule
Parameters & $m=45,  \zeta=-1.035 \times 10^{-7}$  \\
\midrule
$H$ &
\includegraphics[width=5cm]{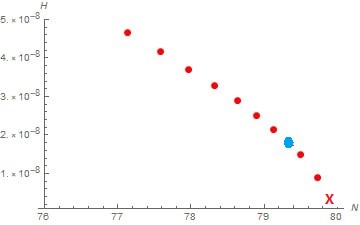} \\
\midrule
$\ddot{a}$ & \includegraphics[width=5cm]{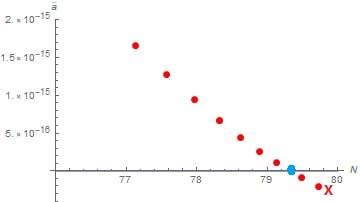} \\
\bottomrule
\end{tabular}
\label{Transient}
\caption{Transition from accelerated into decelerated phases, assuming $H=m/t + \zeta$ .}
\end{table}

  In order for the quintessence field $\phi$ generating the inflation to contribute to the whole historical evolution of the universe including the dark energy era justifying the current accelerating expansion, the perturbation of Eq.  (\ref{eq:Hwithexit}) should be looked at differently as a generalization of the power-law expansion (Eq. \ref{eq:PowerLaw_H}), where rather than to have a fixed power ($m$), we consider it now as time dependent:
\begin{equation}
H=\frac{m(t)}{t}
\end{equation}

As seen above, with $m(t)=m_0 + \zeta t$ one can account for a transition into a decelerated phase, but without being capable of justifying the current accelerating expansion. Looking at Eq. (\ref{eq:Hwithexit}) as a first-order approximation for the function $m(t)$, one can conceive more general forms of $m(t)$. For instance, with $m(t)=m_0  + \zeta t + \xi t^2$ and assuming $\xi > \frac{\zeta^2}{4 m_0}$ then the expanding state ($\dot{a} > 0$) is enforced. For this quadratic form in $t$, one finds:
\[ \dot{H}+H^2 = \frac{m_0^2-m_0}{t^2} + \frac{2m_0 \zeta}{t} + \zeta^2 + 2 \xi m_0 +\xi + 2 \xi \zeta t + \xi^2 t^2
\] and thus with $m_0 \geq 1$ we assure that $\ddot{a}$ is positive for early and late times. With suitable choice of $\zeta$ and $\xi$ one  can find situations with three phases as in the figure table (\ref{TransientQuadratic}). However, and as discussed before, this idea needs a deeper investigation, as the figure here, albeit successful qualitatively in predicting three phases, is still misleading quantitatively in that the regime of $\ddot a <0$ is just a short transient (of the order of $10^6$ Planck times, i.e $\mathcal{O} (10^{-45})$ years) far shorter than actual non accelerating expansion era. This can be better elucidated when represented in terms of the post-inflation efolds increase $\Delta N=\int_{t_e} H(s) ds$ instead of time $t$ (the bottom part of the figure table), where, for the taken values of $m_0=2, \zeta=-2\times 10^{-6}$ and $\xi=5.1\times10^{-13}$, the scenario can only effectively stop the accelerated inflationary regime
by around one efold, whereas a truly graceful exit scenario
should have led to around 60 efolds of non accelerated expansion after inflation.

\begin{table*}[ht]
\begin{tabular}{cc}
\toprule
Parameters & $m_0=2, \zeta=-2\times 10^{-6}, \xi=5.1\times10^{-13}$  \\
\midrule
\includegraphics[width=0.5\textwidth]{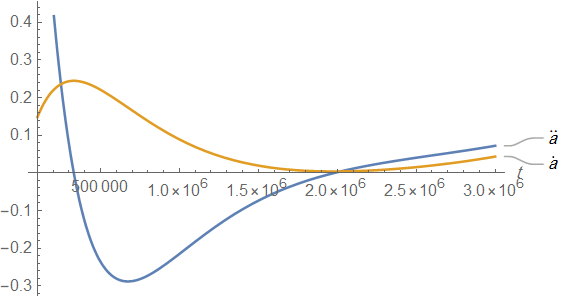} &
\includegraphics[width=0.5\textwidth]{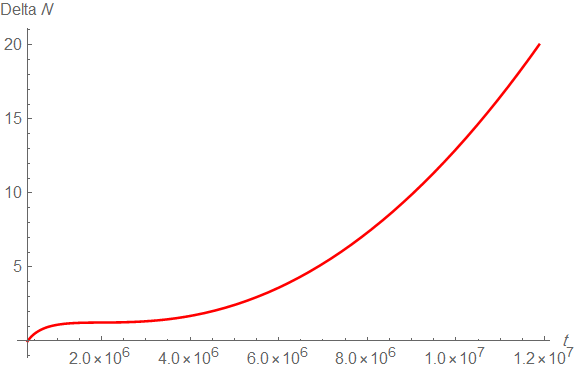} \\
\midrule
\includegraphics[width=0.5\textwidth]{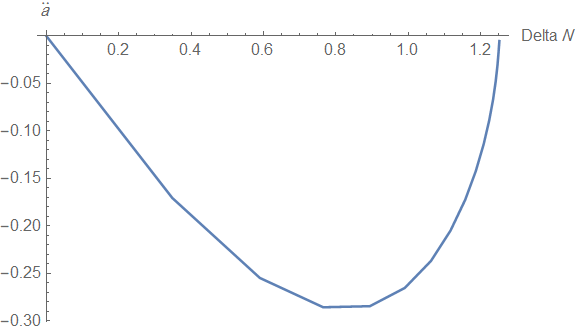} &
\includegraphics[width=0.5\textwidth]{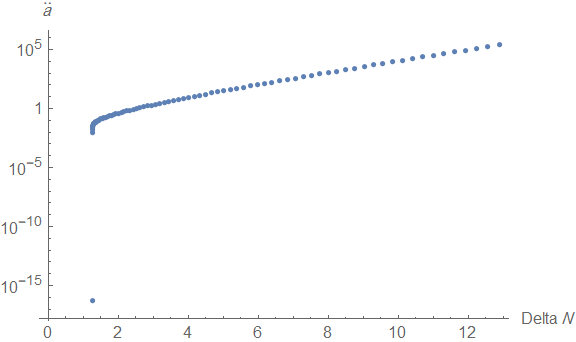} \\
\bottomrule
\end{tabular}
\caption{Top Left: plot of the acceleration $\ddot{a}$  and velocity $\dot{a}/10^6$ rates of cosmic expansion as a function of physical time $t$ in the case of $m(t)=m_0+\zeta t+\xi t^2$, with $m_0=2$, $\zeta=-2\times 10^{-6}$ and $\xi=5.1\times10^{-13}$. Top Right: the post-inflation e-folding increase $\Delta N (t)=\int_{t_e}^t H(s) ds$ in terms of $t$. Bottom: The acceleration rate $\ddot{a}$  in terms of the efolding increase $\Delta N$ after inflation, where the left part corresponds to non accelerated expanding universe.\label{TransientQuadratic}}
\end{table*}

Finally, applying the ``perturbed" modification of Eq. (\ref{eq:Hwithexit}) and adopting the potential of Eq. (\ref{eq:U})  within the cold scenario can improve the latter's results. This improvement is illustrated in Fig. (\ref{fig:ns_r_weak_warm}), where the observables for the cold power-law inflation are shown as magenta dashed lines. The figure shows a satisfactory improvement in the results, which meet the $68\%$ confidence level for Planck 2018 results; otherwise, it is still far from the planck+ BK18 contours.

\section{Summary and Conclusion}
In this work, we studied power-law inflation in the context of a warm inflation scenario. Three dissipation parameters are considered: $\Gamma=\Gamma_0$, $\Gamma=\Gamma_0 T$, and $\Gamma=\Gamma_0 T^3$, with weak and strong dissipation rates.

Similarly to cold inflation, we found that the potential causing the power-law expansion in the weak dissipation limit is exponential. However, whereas the cold scenario gives a too large $r\sim 0.3$ making it unviable, we find that ($n_s, r$) in the weak regime, for all the various cases according to the dependence of $\Gamma$ on temperature, agree largely with the observational results. Still, the weak regime does not have an exit scenario, like the cold paradigm. Considering the dynamics of $Q$, we see that only a part of the exponential potential is valid for the weak limit regime, however for the third case of cubically $T$-dependent $\Gamma$ the entirety of the potential is consistent with the weak regime.

A strong dissipation limit was then applied to the model. As opposed to the weak case, we observed different analytical forms for the potential according to the different $\Gamma(T)$ cases. The study  showed that the model is excluded in both cases of $\Gamma=\Gamma_0$ or $\Gamma=\Gamma_0 T$, since they lead to spectral indices greater than or equal to one. The $\Gamma=\Gamma_0 T^3$ case, however, showed excellent concurrence with observations. Still, the strong regime does not either have an exit scenario because the slow-roll parameters remain constant and independent of the field. Dynamics of $Q$ shows that only a portion of the potential is consistent with the strong regime.

Furthermore, we proposed a graceful scenario assuming a small Hubble parameter perturbation. Based on this modification, we examined the weak dissipation limit, since -unlike the strong limit- analytical expressions for the potential could be obtained. The corresponding `perturbed' potential is a sum of three terms, one of which can play the role of cosmological constant in the large field limit. The `perturbed' spectral observables ($n_s, r$) are very close to their `unperturbed' counterparts, and the model is viable in this respect. Moreover, unlike the pure unperturbed case, one can reach now the end of inflation in both cases of ($\Gamma = \Gamma_0$ and $\Gamma = \Gamma_0 T$) lying consistently before reaching the boundary of the weak limit regime, and enough e-folds have been produced to overcome the flatness problem of the big bang theory. For the case ($\Gamma = \Gamma_0 T^3$), and unlike the unperturbed scenario, only one portion of the potential is consistent with the weak limit regime, but no consistent end of inflation within this weak limit portion.

Finally, by applying the same `perturbed' modification to the cold power law inflation, we found an indicative way for improvement, when compared
with the pure unperturbed case, towards a possible graceful exit.

\begin{acknowledgements}
We thank E. I. Lashin for useful discussions. N. Chamoun acknowledges support from PIFI-CAS program, from Humboldt Foundation and from ICTP Associate program. The authors are grateful to the President of Damascus University for his support, and to the anonymous referee for critical comments.
\end{acknowledgements}



\end{document}